\documentclass[submission,copyright,creativecommons]{eptcs}
\providecommand{\event}{AiSoS 2013} 
\usepackage{mathptmx}
\usepackage{helvet}
\usepackage{courier}
\usepackage[T1]{fontenc}
\usepackage[latin9]{inputenc}
\usepackage{graphicx}
\usepackage[numbers]{natbib}

\usepackage{marginnote}

\let\myMarginpar\marginpar\renewcommand{\marginpar}[1]{\myMarginpar{\hspace{0pt}\scriptsize{\\#1}}}

\title{Variability and Evolution in Systems of Systems}
\author{Goetz Botterweck
\institute{Lero--The Irish Software Engineering Research Centre,\\
Limerick, Ireland}
\email{goetz.botterweck@lero.ie}
}
\def\titlerunning{Variability and Evolution in Systems of Systems}
\def\authorrunning{Goetz Botterweck}
\begin{document}
\maketitle

\begin{abstract}
In this position paper (1) we discuss two particular aspects of Systems
of Systems, i.e., \emph{variability }and \emph{evolution}. (2) We
argue that concepts from Product Line Engineering and Software Evolution
are relevant to Systems of Systems Engineering. (3) Conversely, concepts
from Systems of Systems Engineering can be helpful in Product Line
Engineering and Software Evolution. Hence, we argue that an exchange
of concepts between the disciplines would be beneficial.
\end{abstract}

\section{Introduction}

In this position paper we (1) discuss two particular aspects of Systems
of Systems (SoS), i.e., \emph{variability }and \emph{evolution}. We
do this from two perspectives: (2) First, we argue that in order to
address variability and evolution in the context of Systems of Systems
Engineering (SoSE), concepts from Product Line Engineering (PLE) and
Software Evolution are relevant and helpful. (3) Second, we observe
that with increasing maturity of the disciplines the ``objects of
engineering'' in PLE and Software Evolution become larger and larger.
Consequently, such disciplines have to deal with SoS challenges. Hence,
in this paper, we suggest a more lively exchange of concepts between
the disciplines. We are aware that this paper mostly raises questions
and does not provide a lot of answers. However, we strongly believe
that such an exchange would be interesting and beneficial for both
communities.

\begin{figure}
\begin{centering}
\includegraphics[width=0.9\textwidth]{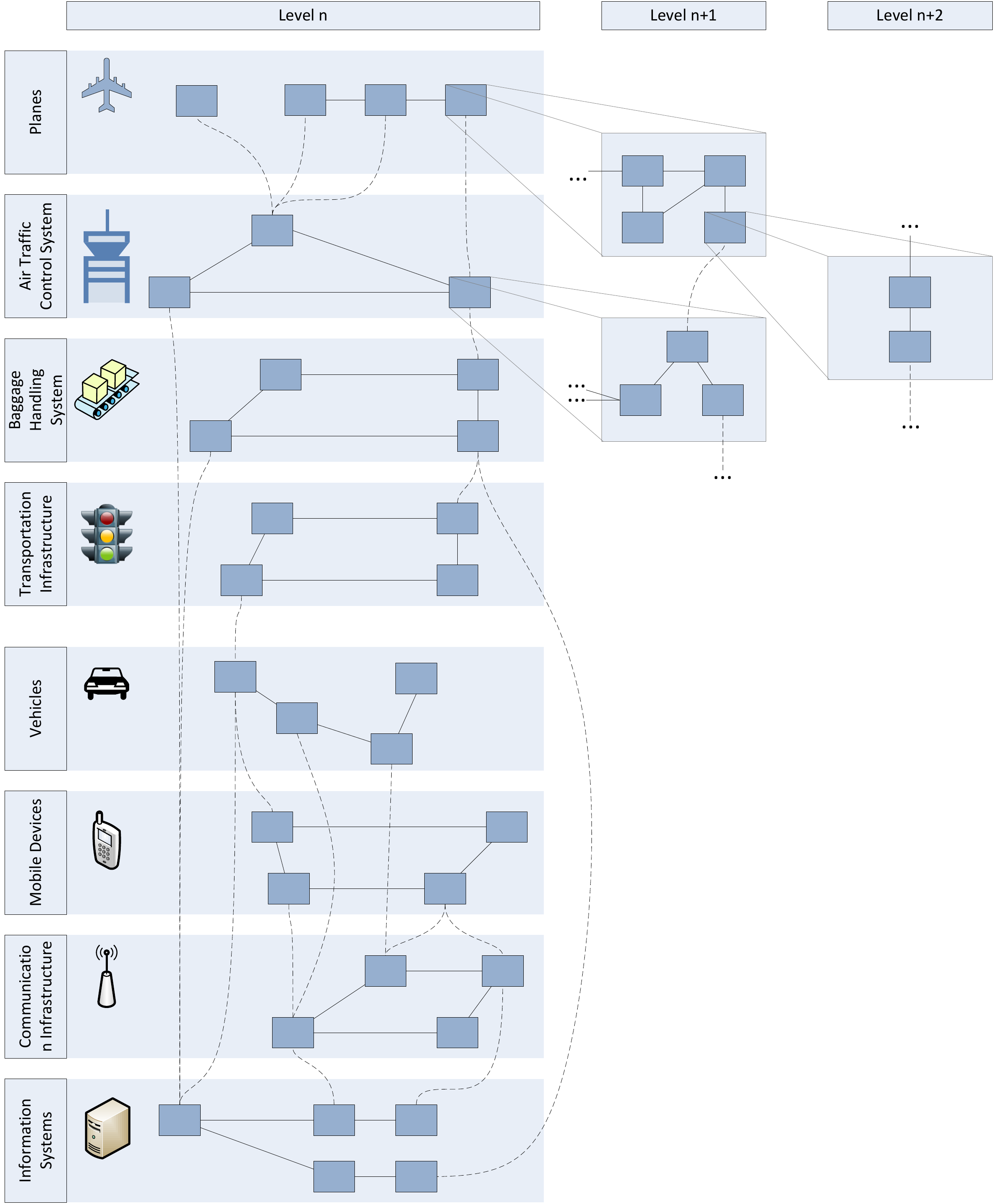}
\par\end{centering}

\caption{An airport as an example of a system of systems; including the recursive
composition out of subsystems\label{fig:airport}}
\end{figure}

As an running example for the further discussion consider the scenario
of an airport, schematically illustrated in Figure~\ref{fig:airport}.
Vertically, we distinguish \emph{subsystems} like Planes, Air Traffic
Control System, Baggage Handling System, Transportation Infrastructure,
etc. Please note that the vertical presentation serves only to distinguish
the subsystems and is \emph{not} intended to indicate a layered model
like for instance used in networking architectures.

Each subsystem consists of \emph{elements} which can be in turn systems
in their own right~\citep{Maier1998Architecting}. This hierarchical
compositional structure can occur recursively over many levels. For
instance, the Planes subsystem consists of planes; a plane consists
of cockpit, navigation subsystem, power subsystem etc.; the power
subsystem consists of engines, fuel supply, etc. Elements are connected
to each other, either within the subsystems (solid lines) or across
subsystem boundaries (dashed lines).

In such systems we have to deal with \emph{variability} and \emph{evolution}.
First, in many cases it is beneficial to consider multiple elements
together (e.g., when designing and implementing them) rather than
addressing each element individually~\citep{Parnas1976Design}. For
instance, when creating and maintaining the communication infrastructure
subsystem and its elements, it is beneficial to consider \emph{types}
or \emph{families }of network routers rather than discussing each
router individually. We then have to deal with variability and commonality
among these elements. We will discuss this aspect in detail in Section~\ref{sec:Variability-in-Systems-of-Systems}.

Second, we have to consider how systems and their elements evolve
over time. Long-term evolution is an inherent characteristic of Systems
of Systems. Even though evolution in such large systems cannot be
controlled by one particular party, we are nevertheless interested
in systematic approaches to evolution. We will discuss this aspect
of evolution in Section~\ref{sec:Evolution-in-Systems-of-Systems}.

We will now give a brief overview of background information (Section~\ref{sec:Background}).
Then, we will look at the aspects of variability and evolution in
SoS (Sections~\ref{sec:Variability-in-Systems-of-Systems} and \ref{sec:Evolution-in-Systems-of-Systems}).
The paper concludes with final thoughts (Section~\ref{sec:Conclusions}).

\section{Background\label{sec:Background}}

In this section we will give a brief summary of relevant background
concepts from the disciplines Systems of Systems Engineering, Product
Line Engineering, and Software Evolutions. Readers that are familiar
with an area might skip over that particular section.

\subsection{Systems of Systems Engineering}

In this paper, we are discussing the relationship between Systems
of Systems Engineering, Product Line Engineering, and Software Evolution.
For instance, we are interested in how the inherent properties of
SoS influence other engineering practices. Hence, let us first look
at these defining properties of SoS.

Systems of Systems have been widely discussed in the literature (e.g.,
\citep{Maier1998Architecting,Cook2001acquisition,Sage2001systems}).
Even though no universally accepted definition exists, some \textbf{characteristics}
have been described, e.g., by Maier~\citep{Maier1998Architecting}:
\begin{itemize}
\item Operational Independence of Elements
\item Managerial Independence of Elements
\item Evolutionary Development
\item Emergent Behavior
\item Geographical Distribution of Elements
\end{itemize}
In addition, we can observe the following phenomena which raise \textbf{additional
challenges} \textbf{for engineering practices} (e.g., Systems Engineering,
Software Engineering, Product Line Engineering) (based on \citep[p.173--174]{Chen2003Advancing}):
\begin{itemize}
\item Multiple stakeholders (for instance, related to the different subsystems)
with varying, potentially conflicting interests
\item High levels of technical complexity
\item Large-scale, broad scope, long-term activity
\item Change management and evolution management are relevant in many aspects
and parts of the system
\item Various constituent systems with independent life-cycles and lines
of responsibility
\item Requirement for adaptability, flexibility and open interfaces
\end{itemize}
Similar to Systems of Systems some authors discuss the concept of
a \emph{Federations of Systems} (e.g., \citep{Sage2001systems}),
which takes the idea even further, e.g., in terms of absence of a
central authority. In this context, Krygiel~\citep{Krygiel1999Behind}
describes a hierarchical taxonomy of conventional systems, systems
of systems (SoS) and federations of systems (FoS) .

\begin{figure}
\begin{centering}
\includegraphics[width=0.85\textwidth]{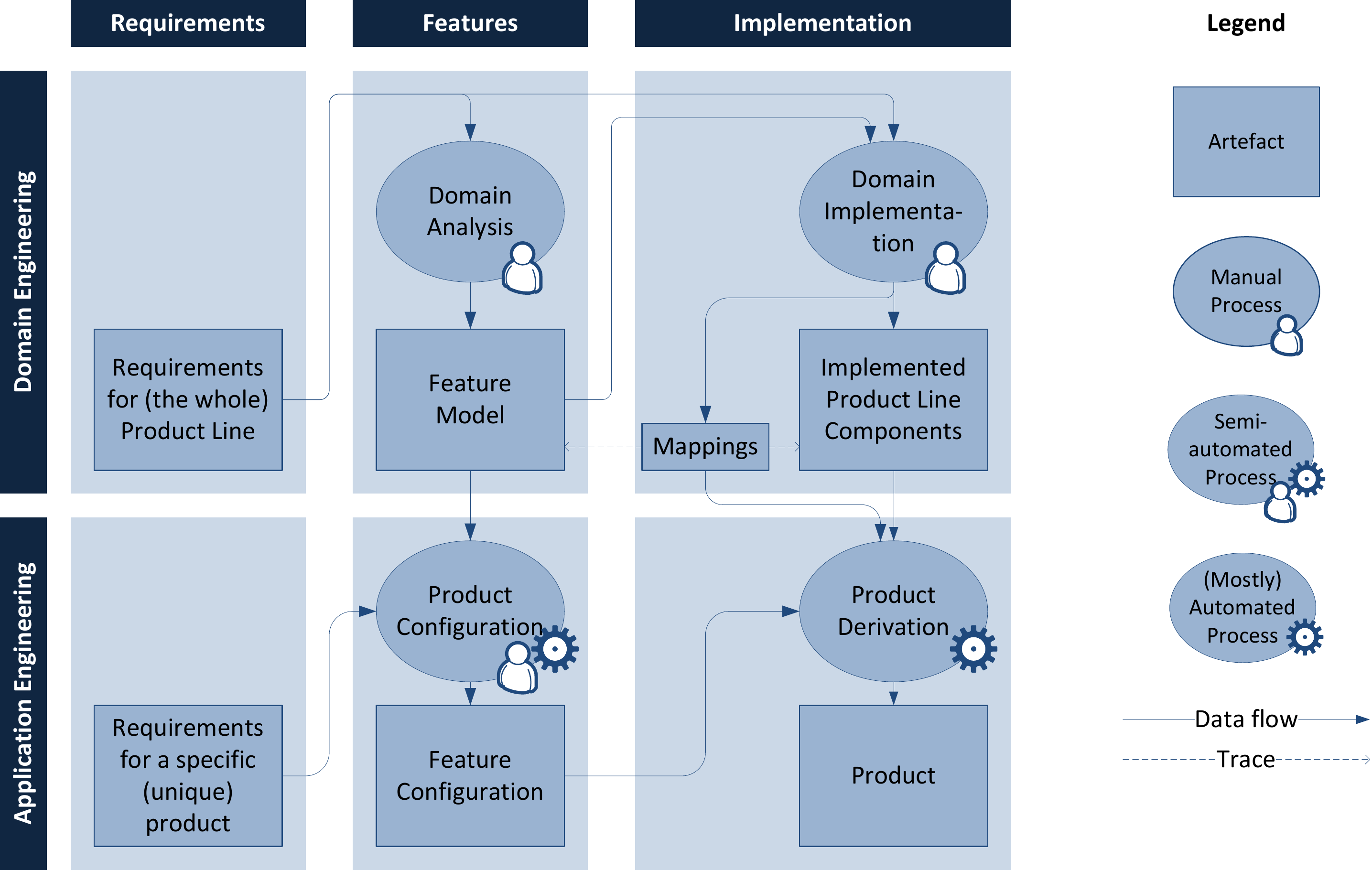}
\par\end{centering}

\caption{Software Product Line Engineering\label{fig:Software-Product-Line-Engineering}}
\end{figure}

\subsection{Software Product Line Engineering}

As a structured approach for variability management and systematic
reuse, we will now look at the discipline of Software Product Line
Engineering. Software Product Line Engineering~\citep{CN2002Software,PBL2005Software}
commonly distinguishes two phases of engineering (Figure~\ref{fig:Software-Product-Line-Engineering}):

In \emph{Domain Engineering }the product line is created (e.g., by
establishing a scope and implementing shared assets). In \emph{Domain
Analysis} the scope of the product line and the allowed variations
among its products are defined. To represent available variants configuration
choices and to define how choosing particular options will influence
the implementation-oriented artifacts, PLE approaches usually define
some form of variability model. A popular group of approach are \emph{Feature
Models} (e.g., \citep{KCH+1990Feature,CE2000Generative}).%
\footnote{Other techniques are \emph{decision models (e.g., \citep{SS2000PuLSE-BEAT,Dhungana2011Dopler}})
or the \emph{orthogonal variability model (OVM)~}\citep{PBL2005Software}.
Surveys of variability modeling approaches can be found in \citep{Chen2009Variability,CGR+2012Cool}.%
} In \emph{Domain Implementation} corresponding assets are created
and mapped to the features to define how a particular configuration
decision will influence the implementation.

In \emph{Application Engineering} products are derived from the platform,
which the product line provides. This process starts with the process
of \emph{Product Configuration} where the user is making configuration
decisions based on the feature model, reducing the set of products,
until exactly one product remains, represented as a \emph{Feature
Configuration}. This process can be supported by interactive tools
(e.g., \citep{Beuche2008Modeling,Botterweck2012S2T2}). In \emph{Product
Derivation} the corresponding implementation is created by composition
and/or generative approaches. Ideally, this step can be mostly automated.

Product Line Engineering can be used on different levels of sophistication.
Bosch~\citep{Bosch2002Maturity} gives a taxonomy of product line
approaches, including a discussion of their maturity. Of the discussed
approaches, \emph{Programme of Product Lines} goes somewhat into the
direction of a System of Systems, but does not quite reach its complexity.
Bosch speaks of very large systems with ``a software architecture
that is defined for the overall system and that specifies the components
that make up the system. {[}..{]} The configuration of the components
is performed {[}..{]} through product line-based product derivation''.

In general, there is little discussion in the PLE literature of products
as SoS, e.g., that they are composed of subsystems or part of a larger
super-system. In particular, it is rarely considered how this hierarchical
structure of systems should be reflected in the various PLE techniques.
We will discuss this in more detail in Section~\ref{sec:Variability-in-Systems-of-Systems}
.

\subsection{Software Evolution}

Another relevant area related to the engineering of complex software-intensive
systems is Software Evolution~\citep{Mens2008Software}. Traditionally,
the literature in Software Evolution has mainly focused on analyzing
\textbf{evolution ``in hindsight}\textbf{\emph{''}}, after it happened.
We could also call this form of evolution \emph{descriptive}, since
it aims to describe how the evolution happened in the past. For instance,
there are Lehman's Laws of Software Evolution~\citep{Lehman1996Laws,Lehman2001Rules}:
\begin{itemize}
\item Continuing Change - a system must continually be adapted to its changing
environment or it will become progressively less satisfactory
\item Increasing Complexity - as a system evolves its complexity increases
unless active work is undertaken to reduce it
\item Self-regulation - the evolution process is self-regulating with close
to normal distribution of measures of product and process attributes
\item Organizational stability (invariant work rate) - the average effective
global activity is invariant over the systems life time
\item Conservation of Familiarity - the content of successive releases is
statistically invariant
\item Continuing Growth - the functional content of the system must continually
be increased to maintain stakeholder satisfaction
\item Declining Quality - the quality of the system will be (perceived as)
declining unless rigorously maintained and adapted to a changing environment
\item Feedback System - evolution processes are multi-level, multi-loop,
multi-agent feedback systems and must be treated as such to be successfully
improved
\end{itemize}
These ``laws'' have been observed and analyzed on numerous cases.
Even though in the discussion of Lehman's laws, the term of ``systems''
is used, the SoS concept is really considered.

In contrast to the analysis of evolution ``in the past'', other
works are focusing on \textbf{planned and managed evolution. }One
could call this form of evolution \emph{prescriptive}, since it aims
to create and form a certain reality through evolution. Some authors
propose apply modeling and model-driven techniques for this (e.g.,
\citep{DeursenVW07}). In our earlier work we combine concepts from
Product Line Engineering and Software Evolution to support the feature-oriented
evolution of product lines~\citep{Pleuss2012Model,SPB+2012Modeling}.

Again, the literature on Software Evolution rarely touches the SoS
aspect. In the other direction, in Systems of Systems Engineering,
however, the aspect of evolution is discussed as one of the challenges.
We will look at this in more detail in Section~\ref{sec:Evolution-in-Systems-of-Systems}.

\section{Variability in Systems of Systems\label{sec:Variability-in-Systems-of-Systems}}

In this section we will look at the interplay between Systems of Systems
and Variability. Correspondingly, we will discuss relationships between
concepts in Systems of Systems Engineering and Product Line Engineering.

\begin{figure}
\begin{centering}
\includegraphics[width=1\textwidth]{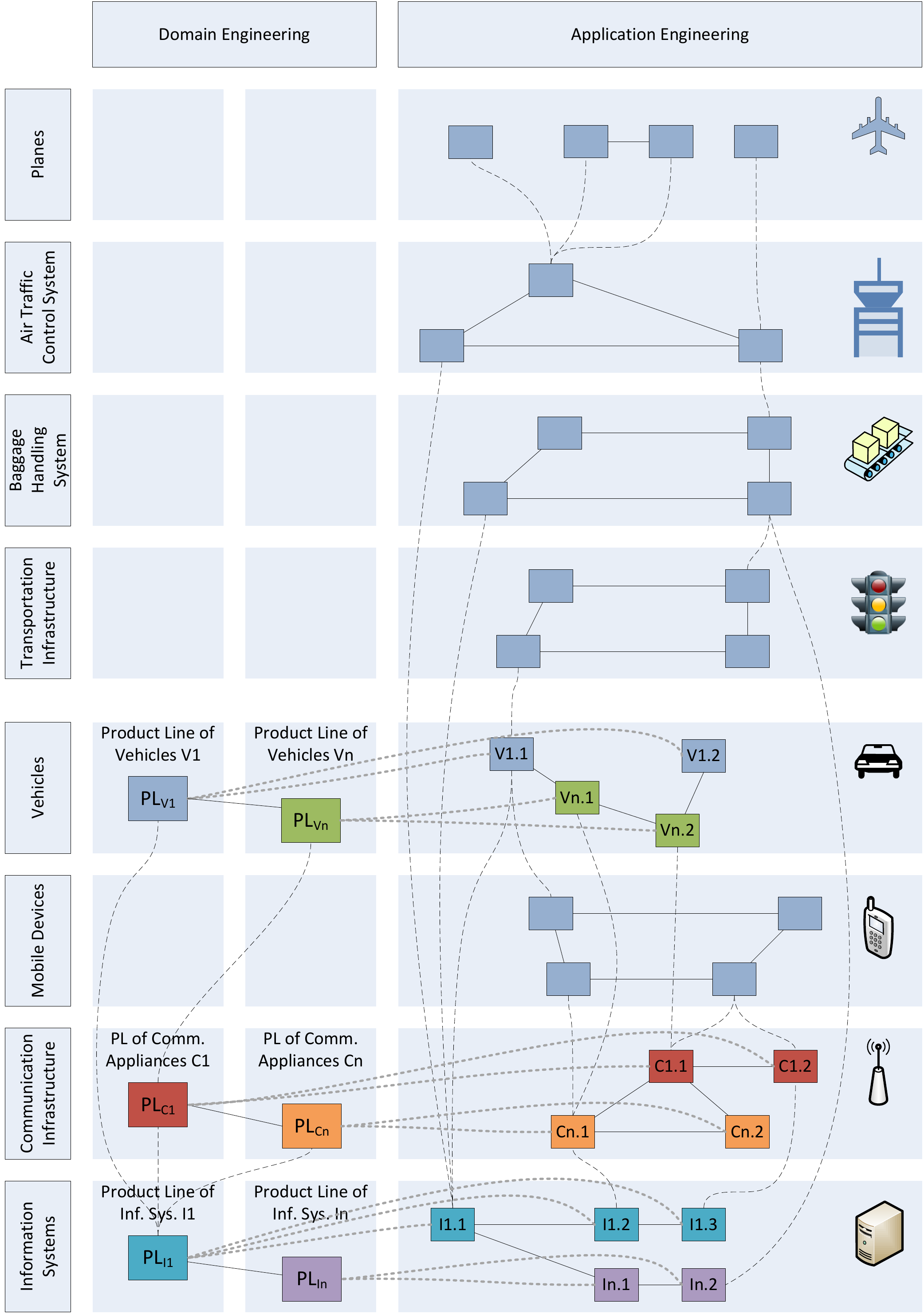}
\par\end{centering}

\caption{Again the airport as a Systems of Systems; here with concepts from
Product Line Engineering and \label{fig:airport-withPLE}}
\end{figure}

\subsection{Structures and their modeling}

First of all, when dealing with Systems of Systems Engineering, we
can consider each system to be potentially a product of a product
line. For illustration purposes see the updated airport scenario in
Figure~\ref{fig:airport-withPLE}. It has been extended to show the
distinction between product lines (Domain Engineering, left-hand side)
and products (Application Engineering, right-hand side). This model
is a combination of the airport scenario (Figure~\ref{fig:airport})
and the PLE framework (Figure~\ref{fig:Software-Product-Line-Engineering}).
Please note that we abstract from details by representing product
lines as whole entities, not distinguishing various artifacts within
the product line. Also here the distinction between Domain Engineering
and Application Engineering is left-vs.-right, whereas in Figure~\ref{fig:Software-Product-Line-Engineering}
it was top-vs.-bottom. For simplification the diagram shows only product
lines for three subsystems. We have for instance , $V1.1$ and $V1.2$
as products of the $PL_{V1}$ product line of vehicles and $Vn.1$
and $Vn.2$ as products of the $PL_{Vn}$ product line.

The motivation to consider systems (in a SoS context) to be products
of a product line can come from multiple drivers. First, in many cases
a \emph{supplier }of systems (e.g., a manufacturer of network routers)
will have families of similar systems and product line techniques
promise considerable benefits in handling such families of products
in a systematic fashion. So product lines can be seen as a technique
to produce components in a SoS approach. Second, from the perspective
of \emph{user }of systems (e.g., the manager of the communication
infrastructure of the airport) it can be beneficial to handle groups
of systems together rather than addressing each system individually.
For instance, the manager could use feature modeling to describe the
variations of network appliances that are currently in operation in
the airport.

As a first implication of the SoS context, we can observe that analogously
to the co-existence and collaboration of multiple systems, we can
expect a \textbf{co-existence of multiple product lines}. Also, structures
between systems have to be reflected between product lines. For instance,
in order to allow a communication link between the vehicle $Vn.2$
and the communication component $CI1.1$ the corresponding product
lines $PL_{Vn}$ and $PL_{CI1}$ must be prepared to provide such
products, e.g., by having corresponding implementations in their asset
base.

Another aspect is that the \textbf{recursive hierarchical composition
}of Systems of Systems can be applied to products as well. For instance,
if we consider a plane as a product (which can be derived from a product
line of planes) then the particular subsystems (e.g., cockpit displays
and controls, navigation subsystem, power subsystem) can be considered
products as well. We then have to consider various relationships,
e.g., the plane \emph{consisting} of cockpit subsystems etc., the
cockpit subsystem \emph{communicating with }the navigation subsystem,
and so on. These relationships have to be reflected between product
lines.

\begin{figure}
\begin{centering}
\includegraphics[width=0.8\textwidth]{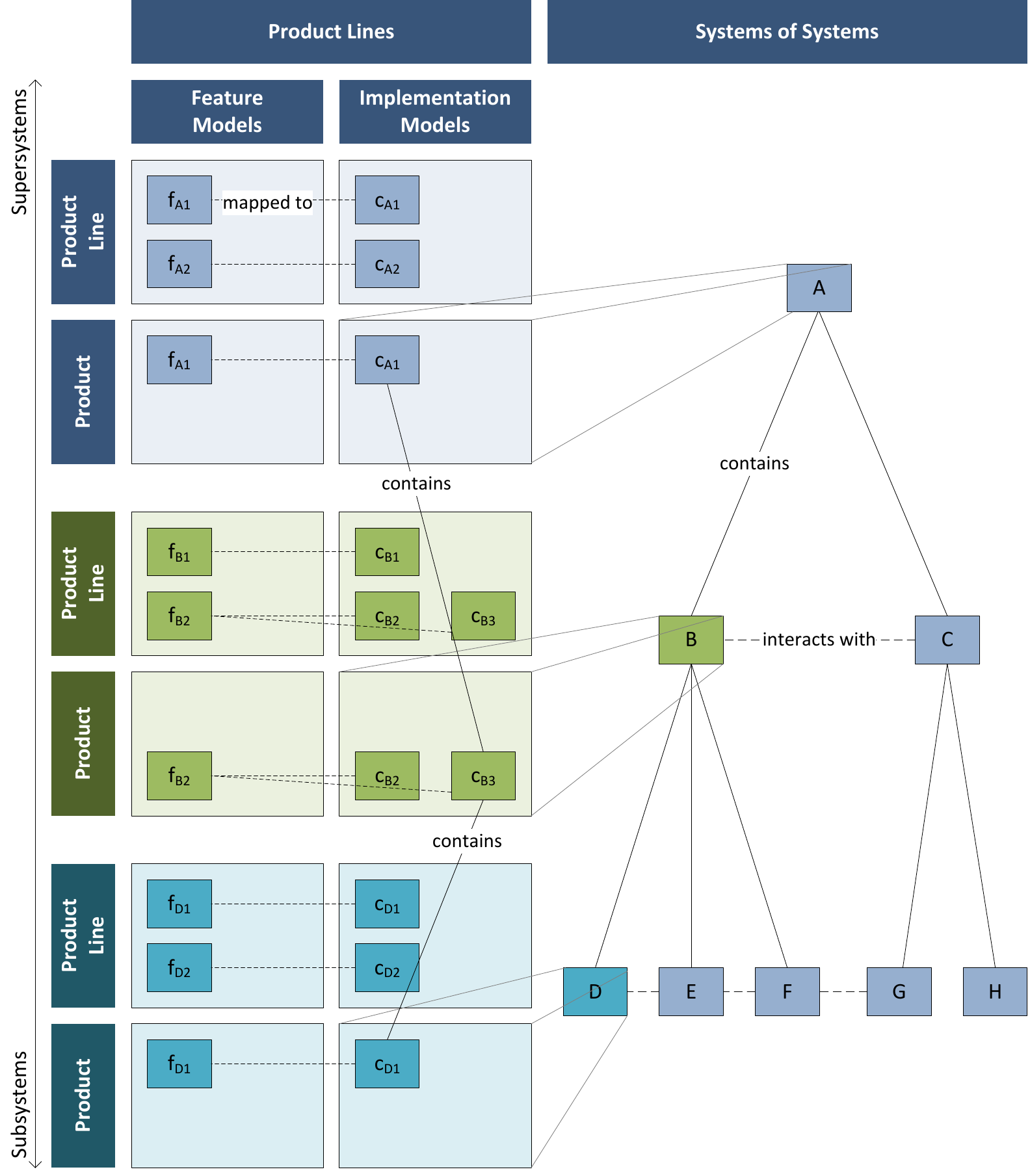}
\par\end{centering}

\caption{Modeling of product lines for systems of systems}
\end{figure}

In the literature there are few approaches in this direction. For
instance, Thompson and Heimdahl~\citep{Thompson2003Structuring}
discuss hierarchical product lines. A related aspect is the modularization
of feature models, e.g., the multilevel feature trees suggested by
Reiser and Weber~\citep{Reiser2006Managing}.

\subsection{Changes and challenges}

We will now consider what consequences are induced by the Systems
of Systems context for Product Line Engineering.

Product Line Engineering has certain assumptions. For instance, many
PLE approaches assume that the domain and scope of the product line
is relatively stable, such that the investment in the product line
can pay off. Also, it is often implicitly assumed that there is \emph{one
}organization that controls how products are built, etc. We need to
consider how these assumptions change when we move to a SoS context
and what challenges arise from that.

\subsubsection{Configuration.}

In product configuration we can observe the following effects of a
System of System context:

First, the \textbf{hierarchical recursive structure} of the SoS needs
to be reflected in the configuration process. For instance, when configuring
a plane, we can consider that this will be part of an airport (actually
multiple airports when traveling), and that it contains various subsystems.
If we configure that the plane will have a certain size and wing design,
then this influences how many engine we require and under which constraints
(e.g., delivered thrust, fuel consumption) they must operate. Potentially,
we have to propagate constraints. Depending on the scenario and the
power to ``dictate'' conditions, this occur in various directions:
\begin{itemize}
\item \emph{System to subsystem} - For instance, when the airport configuration
determines the maximal size of planes that can operate on that airport.
\item \emph{System to supersystem} - For instance, when the large Airbus
A380 ``demands'' that the airport gets extended (in its various
subsystems) in order to attract the potential business.
\item \emph{System to a neighboring system on a similar level} - For instance,
when the Airbus A380 requires that new gates with enough passenger
walkways are constructed. This can be considered as an effect of the
preceding case.
\end{itemize}
Second, the \textbf{disappearance of a central control }and increased
importance of multiple stakeholders with potential conflicting views
needs to be reflected when configuring and deriving a product.

In the literature we find a some approaches that can be considered
to help here. Czarnecki et al.~\citep{Czarnecki2004Staged} present
their approach for Staged Configuration. This could be applied to
make first major decisions (on supersystem level) and then later refine
them (for subsystems). However, the approach does not provide concepts
for structuring large models, e.g., by modularization.

Dhungana et al. ~\citep{Dhungana2008Supporting} present an approach
that deals large variability models that are created by multiple stakeholders,
with fragmentation, the need to merge fragments, and to remove inconsistencies
(e.g., variables that have been named differently by different stakeholders).

\begin{figure}
\begin{centering}
\includegraphics[width=0.75\textwidth]{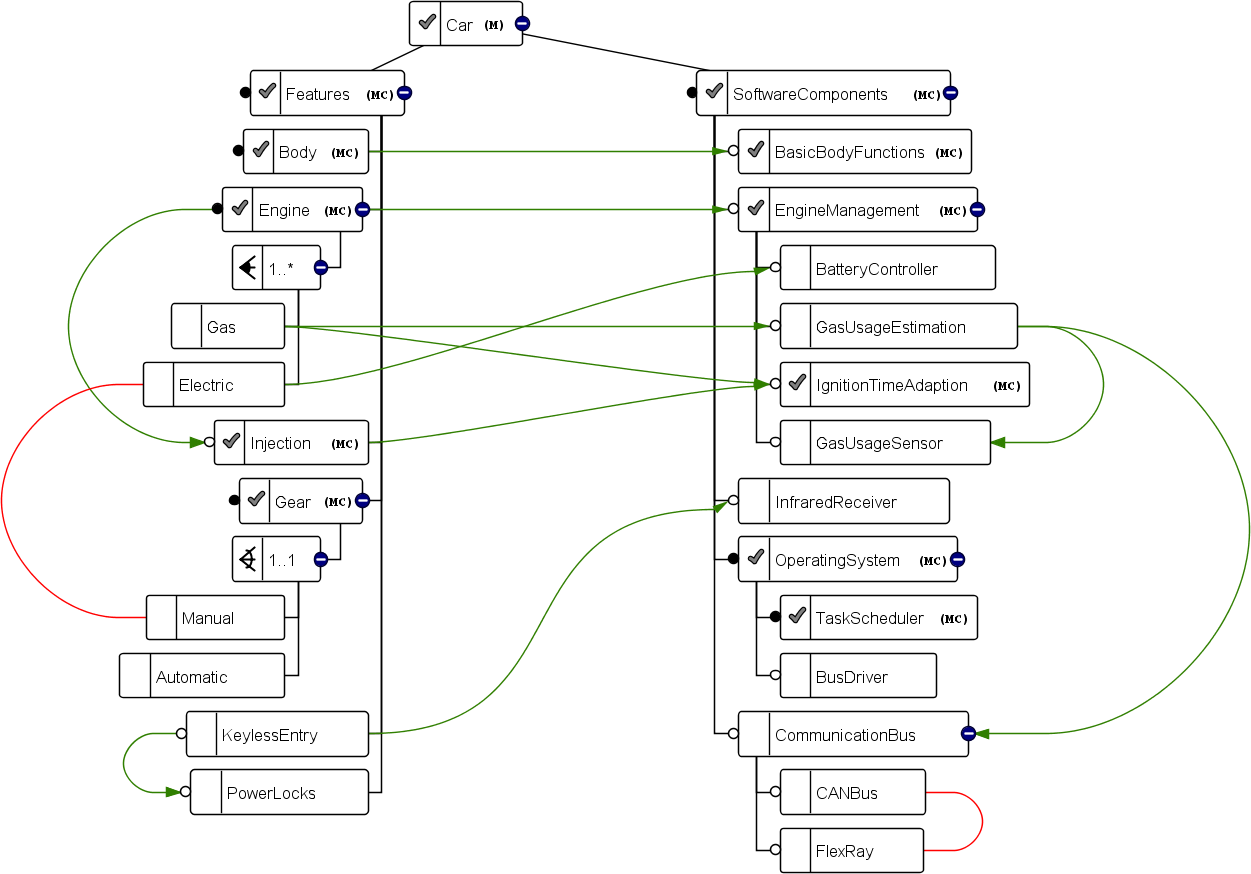}
\par\end{centering}

\caption{Configuring two related feature models side-by-side, taken from~\citep{Pleuss2012Visualization}\label{fig:Configuring-feature-models}}
\end{figure}

In earlier work~\citep{Pleuss2012Visualization}, we have presented
an approach for interactive configuration of feature models, including
interaction techniques that allow to configure two related \textbf{feature
models side-by-side}. See the example in Figure~\ref{fig:Configuring-feature-models}
with a feature model for a car on the left and the corresponding hardware
components on the right. In a SoS context such a configuration tool
could be used to configure two or more related systems side-by-side.

The interactive tool and its reasoning engine calculate and propagate
consequences of configuration decisions. For instance, selecting the
feature \emph{KeylessEntry} (see bottom left) would cause the features
\emph{PowerLocks} and \emph{InfraredReceiver} (in the other feature
model part) to be selected as well. For the SoS discussion it should
be noted that such approaches for modeling of constraints, reasoning
and propagation of consequences often assume that the user who is
performing the configuration (1) has precise \textbf{knowledge} about
all involved systems and (2) the \textbf{power} to actually realize
the chosen options. For instance, what does it help to choose a \emph{KeylessEntry}
function when we cannot ensure that the hardware will actually have
an \emph{InfraredReceiver}?

Dhungana et al.~\citep{Dhungana2011Configuration} present \emph{invar,
}an approach for the joint configuration of heterogeneous variability
models, e.g., to configure a feature model, a decision model, and
an OVM model side-by-side. Such techniques could be extended to support
scenarios, where the ``owners'' of related systems use different
variability modeling approaches.

For product configuration we remaining challenges:
\begin{itemize}
\item Representation and handling of very large models (both in terms of
scale and complexity), including the reflection of the hierarchical,
recursive compositional structure of SoS
\item Dealing with incomplete information (because we might not get all
information about the internals of a neighboring system to which we
interface)
\item Dealing with inconsistent information (because multiple stakeholders
will have conflicting views and decisions)
\end{itemize}

\subsubsection{Analysis.}

There exist several approaches for the analysis of product lines (e.g.,
\citep{BST+2007FAMA,Benavides2010Analysis}). Typical examples for
available analyses is the enumeration of products, the detection of
inconsistencies, or the detection of hidden dependencies (i.e., dependencies
that exist but are not yet modeled explicitly).

When transitioning to an SoS context, analysis techniques face similar
challenges as already discussed for configuration approaches, e.g.,
they have to deal with very large models, the SoS structure, and deal
with incomplete as well as inconsistent information.

\subsubsection{Product Properties.}

Traditionally, product line approaches were often limited to Boolean
concepts and decisions (e.g., a feature is selected or eliminated,
$f_{1}$ requires $f_{2}$). Recently, there has been increased interest
in product attributes and Non-functional Properties (NFP). This includes
the prediction of properties based on a given feature configuration~\citep{Siegmund2012Predicting}
, the configuration (``Give me a plane with top speed over 500 kph'')
and optimization~\citep{Siegmund2011SPL,White2009Selecting} (``Out
of all planes that have my required features, give me the one with
lowest price''). Often we have to deal with soft constraints and
preferences which are often expressed as a utility function (``I
want both low price and high top speed, but low prices is twice as
important''.)

It should be noted that here the PLE community often adapts and applies
techniques that have been published earlier in other fields, e.g.,
in Artificial Intelligence or Constraints (e.g.,\citep{SW1998Product}).

Of particular interest in a SoS context is the article by Siegmund
et al.~\citep{Siegmund2012Interoperability}, which deals with NFP
in complex systems, e.g., a camera-based surveillance system, where
the properties of various parts of the system influence each other
(e.g., resolution of the camera vs. bandwidth of the network).

\subsubsection{Software Architecture.}

System of Systems engineering often deals with architectural frameworks
(e.g., \citep{Carlock2001System}) which are required to give structure
and handle complexity, e.g., by defining how systems can be related
to each other.

Product Line Engineering deals with Product Line Architectures (e.g.,
\citep{Bosch2000Design,Matinlassi2004Comparison}), where a greater
emphasis is on variability and variability implementation. When transitioning
to an SoS context, we have to reflect the structure of systems in
the software architecture. For instance, corresponding to the hierarchical,
recursive compositional structures we can have a nested hierarchy
of product line architectures. The PLA of the supersystem is then
refined by multiple PLAs of its subsystems.

\subsubsection{Organizational structures and processes.}

There are varying organizational structures and process structures
for PLE approaches, e.g., we have consider which organizational units
are responsible for Domain and Application Engineering~\citep{Bosch2002Maturity}.
When transitioning to an SoS context, these organizational structures
have to be re-considered accordingly. For instance, we can have one
Domain Engineering team that is responsible for multiple subsystems
or the responsibility for Domain Engineering can be distributed among
different teams.

\begin{figure}
\begin{centering}
\includegraphics[width=0.7\textwidth]{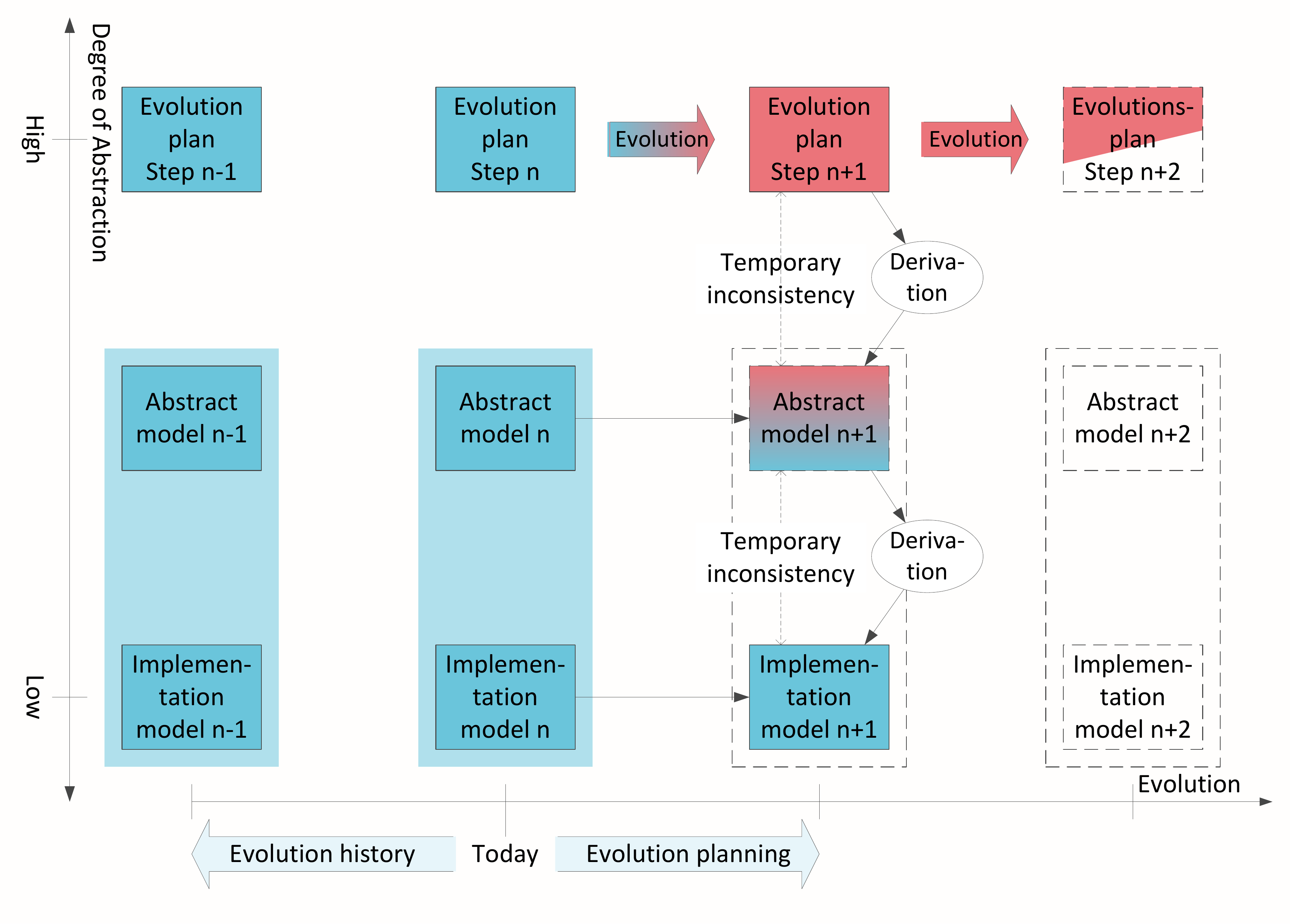}
\par\end{centering}

\caption{Planned evolution~\label{fig:evolution}}
\end{figure}

\section{Evolution in Systems of Systems\label{sec:Evolution-in-Systems-of-Systems}}

In this section we will look at the interplay between Systems of Systems
and Evolution. Correspondingly, we will discuss relationships between
concepts in Systems of Systems Engineering and Software Evolution.

The lack of strategies and techniques for a sustainable evolution
of very large and complex systems is one of the main motivators towards
SoS research.

\subsection{Historic evolution vs. future evolution}

As discussed earlier we can consider software evolution from \textbf{two
perspectives}, the analysis of evolution ``in the past'' and the
planning and management of evolution ``in the future''. In a SoS
context, both aspects are relevant just as well: On the one hand,
we are interested how and where evolution occurs in a SoS, e.g., to
better understand how everything is connected, how changes were propagated,
or to determine if certain ``laws'' hold. On the other hand, we
can aim to plan and manage the evolution happening in a SoS, e.g.,
with techniques to plan evolution over multiple interconnected systems
and techniques to monitor and influence changes while they happen.

For illustration purposes consider Figure~\ref{fig:evolution} horizontally
showing evolution over time from left to right, vertically showing
various abstraction levels including plans for evolution on the topmost
layer. To the left we can look at historic evolution that happened
in the past. To the right we can see planned, future evolution, where
the red annotation indicates an induced change that temporarily causes
inconsistencies and is propagated through the abstraction layers when
these inconsistencies get removed again.

\subsection{Consequences of SoS characteristics for software evolution}

What consequences do Systems of Systems characteristics have for Software
Evolution?

First of all, any approach to software evolution in a SoS context
has to reflect the \textbf{hierarchical, recursive structure of SoS
}and must be able to handle their, scale and complexity. For instance,
when planning and realizing a change in a system, this change has
potentially to be propagated into all subsystems.

Systems of Systems have by their very nature a long life span and
to understand, create, and manage them we have to take a \textbf{long-term
perspective}. Organizations dealing with or operating within Systems
of Systems often have to take this long-term perspective.

There is no obvious reason, why Lehman's laws of evolution~\citep{Lehman1996Laws,Lehman2001Rules}
should not hold for Systems of Systems. For instance, one can expect
that changes in the environment induce necessary changes in the SoS
in order to keep it relevant. Obviously, whether these laws actually
hold needs to be evaluated before more concrete statements can be
made. Moreover, it would be interesting to see how the various variables
and forces in Lehman's laws (e.g., development activity, growth rate,
defect rate) react in a SoS context and how the descriptive models
have to be extended.

Due to the \textbf{managerial independence of elements }and the disappearance
of central control an approach that aims to control evolution for
the whole SoS (e.g., by demanding changes all over the place) is infeasible
in practice, since there is rarely a single entity which has enough
control to implement these changes. Instead, evolutionary objectives
of \textbf{multiple stakeholders} have to be considered and consolidated.
An approach needs to tolerate potentially incomplete and inconsistent
information.

Even though there is an independence of elements, we have to consider
\textbf{dependencies among elements}, when planning and implementing
changes. Also, it is impossible to introduce changes without causing
\textbf{inconsistencies} at least temporarily. Hence, many changes
can only happen incrementally and changes are propagated through an
introduction and subsequent resolution of inconsistencies.

On a related notion, System of Systems undergo \textbf{evolutionary
changes}, which can be seen from two sides: First, change is inevitable
and omnipresent. If systems have to evolve to adapt to changes in
their environment and to stay relevant~\citep{Lehman1996Laws}, the
same can be argued for systems of systems. Second, because of the
size and complexity of Systems of Systems change can only happen in
an evolutionary fashion, not as a disruptive ``flick of a switch''
event. As an example in our airport scenario consider the deployment
of a faster Wi-Fi network, to be used by both personnel and passengers.
This causes changes in several subsystems, e.g., new authentication
mechanisms, updates to the user database, mobile devices supporting
the new standard, changes to software applications that using the
network, etc.

On a process level the dependencies among elements and the evolutionary
nature of changes has to be reflected in the corresponding \textbf{life-cycles}
and processes. The overall ``life-cycle'' of the System of Systems
is composed out of interconnected, interwoven smaller evolution steps
of the individual subsystems and their elements.

\subsection{More of the same?}

Here, one could say that in order to handle evolution of Systems of
Systems, we can just apply well known techniques for Software Evolution,
just more of it. However, we would argue that above a certain level
additional approaches are required.

For instance, Chen and Clothier~\citep{Chen2003Advancing} point
out that the high-level engineering complexity (as discussed further
by \citep{Sage2001systems,Carlock2001System}) raise great challenges
in evolutionary development of SoS and indicates a need for considering
different SE strategies at a level above individual projects.

Some part of SoS evolution can be covered by evolution on a component
level, considering the component and its evolution in isolation. However,
that is not sufficient. The real challenges lie in the coordinated
evolution of multiple systems. Compare our earlier example of deploying
a new network at an airport. Also see the discussion of ``joint evolution''
in \citep[p. 173]{Chen2003Advancing}.

\section{Conclusions\label{sec:Conclusions}}

In this paper, we focused on the aspects of variability and evolution
in a Systems of Systems context. We did so from two perspectives:
First, we argue that Systems of Systems Engineering has to deal with
variability and evolution and, hence, concepts from Product Line Engineering
and Software Evolution can be helpful. Second, these disciplines,
PLE and Software Evolution, more and more have to deal with very large
systems where concepts from Systems of Systems Engineering can be
helpful. We are aware that this paper mostly raises questions and
does not provide a lot of answers. However, we strongly believe that
an exchange between the fields of Systems of Systems Engineering,
Product Line Engineering, and Software Evolution would be interesting
and beneficial for both communities.

\bibliographystyle{eptcs}
\bibliography{sos-selected}

\begin{thebibliography}{10}
\providecommand{\bibitemdeclare}[2]{}
\providecommand{\surnamestart}{}
\providecommand{\surnameend}{}
\providecommand{\urlprefix}{Available at }
\providecommand{\url}[1]{\texttt{#1}}
\providecommand{\href}[2]{\texttt{#2}}
\providecommand{\urlalt}[2]{\href{#1}{#2}}
\providecommand{\doi}[1]{doi:\urlalt{http://dx.doi.org/#1}{#1}}
\providecommand{\bibinfo}[2]{#2}

\bibitemdeclare{inproceedings}{BST+2007FAMA}
\bibitem{BST+2007FAMA}
\bibinfo{author}{D.~\surnamestart Benavides\surnameend},
  \bibinfo{author}{S.~\surnamestart Segura\surnameend},
  \bibinfo{author}{P.~\surnamestart Trinidad\surnameend} \&
  \bibinfo{author}{A.~\surnamestart Ruiz-Cort\'{e}s\surnameend}
  (\bibinfo{year}{2007}): \emph{\bibinfo{title}{{FAMA}: Tooling a Framework for
  the Automated Analysis of Feature Models}}.
\newblock In: {\sl \bibinfo{booktitle}{Proceeding of the First International
  Workshop on Variability Modelling of Software-intensive Systems (VAMOS)}},
  \doi{10.1.1.77.8501}.

\bibitemdeclare{article}{Benavides2010Analysis}
\bibitem{Benavides2010Analysis}
\bibinfo{author}{David \surnamestart Benavides\surnameend},
  \bibinfo{author}{Sergio \surnamestart Segura\surnameend} \&
  \bibinfo{author}{Antonio \surnamestart Ruiz-Cort\'{e}s\surnameend}
  (\bibinfo{year}{2010}): \emph{\bibinfo{title}{Automated analysis of feature
  models 20 years later}}.
\newblock {\sl \bibinfo{journal}{Information Systems}}
  \bibinfo{volume}{35}(\bibinfo{number}{6}), pp. \bibinfo{pages}{615--636},
  \doi{10.1016/j.is.2010.01.001}.

\bibitemdeclare{inproceedings}{Beuche2008Modeling}
\bibitem{Beuche2008Modeling}
\bibinfo{author}{Danilo \surnamestart Beuche\surnameend}
  (\bibinfo{year}{2008}): \emph{\bibinfo{title}{Modeling and Building Software
  Product Lines with Pure::Variants}}.
\newblock In: {\sl \bibinfo{booktitle}{12th International Software Product Line
  Conference (SPLC 2008)}}, \bibinfo{address}{Limerick, Ireland},
  \doi{10.1109/SPLC.2008.53}.

\bibitemdeclare{book}{Bosch2000Design}
\bibitem{Bosch2000Design}
\bibinfo{author}{Jan \surnamestart Bosch\surnameend} (\bibinfo{year}{2000}):
  \emph{\bibinfo{title}{Design and Use of Software Architectures: Adopting and
  Evolving a Product-Line Approach}}.
\newblock \bibinfo{publisher}{Addison-Wesley}, \doi{10.1.1.107.7212}.

\bibitemdeclare{inproceedings}{Bosch2002Maturity}
\bibitem{Bosch2002Maturity}
\bibinfo{author}{Jan \surnamestart Bosch\surnameend} (\bibinfo{year}{2002}):
  \emph{\bibinfo{title}{Maturity and Evolution in Software Product Lines:
  Approaches, Artefacts, and Organization}}.
\newblock In \bibinfo{editor}{Garry \surnamestart Chastek\surnameend}, editor:
  {\sl \bibinfo{booktitle}{Proceedings of the Second Software Product Line
  Conference}}, \bibinfo{series}{LNCS 2379}, \bibinfo{publisher}{Springer},
  \bibinfo{address}{San Diego, CA}, pp. \bibinfo{pages}{257--271},
  \doi{10.1.1.92.3163}.

\bibitemdeclare{inproceedings}{Botterweck2012S2T2}
\bibitem{Botterweck2012S2T2}
\bibinfo{author}{Goetz \surnamestart Botterweck\surnameend} \&
  \bibinfo{author}{Andreas \surnamestart Pleuss\surnameend}
  (\bibinfo{year}{2012}): \emph{\bibinfo{title}{S2T2-Configurator: Interactive
  Support for Configuration of Large Feature Models}}.
\newblock In: {\sl \bibinfo{booktitle}{8th European Conference on Modelling
  Foundations and Applications (ECMFA 2012) Tools Track}},
  \bibinfo{address}{Kgs. Lyngby, Denmark}.
\newblock \urlprefix\url{http://hdl.handle.net/10344/2586}.

\bibitemdeclare{article}{Carlock2001System}
\bibitem{Carlock2001System}
\bibinfo{author}{P.G. \surnamestart Carlock\surnameend} \&
  \bibinfo{author}{R.E. \surnamestart Fenton\surnameend}
  (\bibinfo{year}{2001}): \emph{\bibinfo{title}{System of Systems (SoS)
  enterprise systems engineering for information-intensive organizations}}.
\newblock {\sl \bibinfo{journal}{Systems Engineering}}
  \bibinfo{volume}{4}(\bibinfo{number}{4}), pp. \bibinfo{pages}{242--261},
  \doi{10.1002/sys.1021}.

\bibitemdeclare{inproceedings}{Chen2009Variability}
\bibitem{Chen2009Variability}
\bibinfo{author}{Lianping \surnamestart Chen\surnameend},
  \bibinfo{author}{Muhammad \surnamestart Ali~Babar\surnameend} \&
  \bibinfo{author}{Nour \surnamestart Ali\surnameend} (\bibinfo{year}{2009}):
  \emph{\bibinfo{title}{Variability management in software product lines: a
  systematic review}}.
\newblock In: {\sl \bibinfo{booktitle}{Proceedings of the 13th International
  Software Product Line Conference}}, \bibinfo{series}{SPLC '09},
  \bibinfo{publisher}{Carnegie Mellon University},
  \bibinfo{address}{Pittsburgh, PA, USA}, pp. \bibinfo{pages}{81--90}.
\newblock \urlprefix\url{http://dl.acm.org/citation.cfm?id=1753235.1753247}.

\bibitemdeclare{article}{Chen2003Advancing}
\bibitem{Chen2003Advancing}
\bibinfo{author}{P.~\surnamestart Chen\surnameend} \&
  \bibinfo{author}{J.~\surnamestart Clothier\surnameend}
  (\bibinfo{year}{2003}): \emph{\bibinfo{title}{Advancing systems engineering
  for systems-of-systems challenges}}.
\newblock {\sl \bibinfo{journal}{Systems engineering}}
  \bibinfo{volume}{6}(\bibinfo{number}{3}), pp. \bibinfo{pages}{170--183},
  \doi{10.1002/sys.10042}.

\bibitemdeclare{book}{CN2002Software}
\bibitem{CN2002Software}
\bibinfo{author}{Paul \surnamestart Clements\surnameend} \&
  \bibinfo{author}{Linda~M. \surnamestart Northrop\surnameend}
  (\bibinfo{year}{2002}): \emph{\bibinfo{title}{Software Product Lines:
  Practices and Patterns}}.
\newblock \bibinfo{series}{The {SEI} series in software engineering},
  \bibinfo{publisher}{Addison-Wesley}, \bibinfo{address}{Boston, MA, USA}.
\newblock
  \urlprefix\url{http://resources.sei.cmu.edu/library/asset-view.cfm?assetID=3%
0731}.

\bibitemdeclare{inproceedings}{Cook2001acquisition}
\bibitem{Cook2001acquisition}
\bibinfo{author}{S.C. \surnamestart Cook\surnameend} (\bibinfo{year}{2001}):
  \emph{\bibinfo{title}{On the acquisition of systems of systems}}.
\newblock In: {\sl \bibinfo{booktitle}{Proceedings of the 2001 INCOSE
  International Symposium, Melbourne AU}}.
\newblock \bibinfo{note}{ISBN: 0-9720562-0-3}.

\bibitemdeclare{inproceedings}{Czarnecki2004Staged}
\bibitem{Czarnecki2004Staged}
\bibinfo{author}{K.~\surnamestart Czarnecki\surnameend},
  \bibinfo{author}{S.~\surnamestart Helson\surnameend} \& \bibinfo{author}{U.W.
  \surnamestart Eisenecker\surnameend} (\bibinfo{year}{2004}):
  \emph{\bibinfo{title}{Staged configuration using feature models}}.
\newblock In \bibinfo{editor}{R.~\surnamestart Nord\surnameend}, editor: {\sl
  \bibinfo{booktitle}{3rd International Software Product Line Conference (SPLC
  2004)}}, \bibinfo{volume}{LNCS 3154}, \bibinfo{publisher}{Springer Berlin
  Heidelberg}, \bibinfo{address}{Boston, MA, USA}, pp.
  \bibinfo{pages}{266--283}, \doi{10.1007/978-3-540-28630-1\_17}.

\bibitemdeclare{book}{CE2000Generative}
\bibitem{CE2000Generative}
\bibinfo{author}{Krysztof \surnamestart Czarnecki\surnameend} \&
  \bibinfo{author}{Ulrich~W. \surnamestart Eisenecker\surnameend}
  (\bibinfo{year}{2000}): \emph{\bibinfo{title}{Generative Programming}}.
\newblock \bibinfo{publisher}{Addison Wesley}, \bibinfo{address}{Reading, MA,
  USA}.

\bibitemdeclare{inproceedings}{CGR+2012Cool}
\bibitem{CGR+2012Cool}
\bibinfo{author}{Krzysztof \surnamestart Czarnecki\surnameend},
  \bibinfo{author}{Paul \surnamestart Gr\"{u}nbacher\surnameend},
  \bibinfo{author}{Rick \surnamestart Rabiser\surnameend},
  \bibinfo{author}{Klaus \surnamestart Schmid\surnameend} \&
  \bibinfo{author}{Andrzej \surnamestart W\k{a}sowski\surnameend}
  (\bibinfo{year}{2012}): \emph{\bibinfo{title}{Cool features and tough
  decisions: a comparison of variability modeling approaches}}.
\newblock In: {\sl \bibinfo{booktitle}{Proceedings of the Sixth International
  Workshop on Variability Modeling of Software-Intensive Systems}},
  \bibinfo{series}{VaMoS '12}, \bibinfo{publisher}{ACM}, \bibinfo{address}{New
  York, NY, USA}, pp. \bibinfo{pages}{173--182}, \doi{10.1145/2110147.2110167}.

\bibitemdeclare{inproceedings}{DeursenVW07}
\bibitem{DeursenVW07}
\bibinfo{author}{A.~\surnamestart van Deursen\surnameend},
  \bibinfo{author}{E.~\surnamestart Visser\surnameend} \&
  \bibinfo{author}{J.~\surnamestart Warmer\surnameend} (\bibinfo{year}{2007}):
  \emph{\bibinfo{title}{Model-Driven Software Evolution: A Research Agenda}}.
\newblock In \bibinfo{editor}{Dalila \surnamestart Tamzalit\surnameend},
  editor: {\sl \bibinfo{booktitle}{Proceedings 1st International Workshop on
  Model-Driven Software Evolution (MoDSE)}}, \bibinfo{publisher}{University of
  Nantes}, pp. \bibinfo{pages}{41--49}.
\newblock
  \urlprefix\url{http://swerl.tudelft.nl/twiki/pub/Main/TechnicalReports/TUD-S%
ERG-2007-006.pdf}.

\bibitemdeclare{article}{Dhungana2011Dopler}
\bibitem{Dhungana2011Dopler}
\bibinfo{author}{D.~\surnamestart Dhungana\surnameend},
  \bibinfo{author}{P.~\surnamestart Grünbacher\surnameend} \&
  \bibinfo{author}{R.~\surnamestart Rabiser\surnameend} (\bibinfo{year}{2011}):
  \emph{\bibinfo{title}{The DOPLER Meta-Tool for Decision-Oriented Variability
  Modeling: A Multiple Case Study}}.
\newblock {\sl \bibinfo{journal}{Automated Software Engineering}}
  \bibinfo{volume}{18}(\bibinfo{number}{1}), pp. \bibinfo{pages}{77--114},
  \doi{10.1007/s10515-010-0076-6}.

\bibitemdeclare{inproceedings}{Dhungana2008Supporting}
\bibitem{Dhungana2008Supporting}
\bibinfo{author}{Deepak \surnamestart Dhungana\surnameend},
  \bibinfo{author}{Thomas \surnamestart Neumayer\surnameend},
  \bibinfo{author}{Paul \surnamestart Gr{\"u}nbacher\surnameend} \&
  \bibinfo{author}{Rick \surnamestart Rabiser\surnameend}
  (\bibinfo{year}{2008}): \emph{\bibinfo{title}{Supporting Evolution in
  Model-Based Product Line Engineering}}.
\newblock In: {\sl \bibinfo{booktitle}{12th International Software Product Line
  Conference (SPLC 2008)}}, pp. \bibinfo{pages}{319--328},
  \doi{10.1109/SPLC.2008.26}.

\bibitemdeclare{inproceedings}{Dhungana2011Configuration}
\bibitem{Dhungana2011Configuration}
\bibinfo{author}{Deepak \surnamestart Dhungana\surnameend},
  \bibinfo{author}{Dominik \surnamestart Seichter\surnameend},
  \bibinfo{author}{Goetz \surnamestart Botterweck\surnameend},
  \bibinfo{author}{Rick \surnamestart Rabiser\surnameend},
  \bibinfo{author}{Paul \surnamestart Gruenbacher\surnameend},
  \bibinfo{author}{David \surnamestart Benavides\surnameend} \&
  \bibinfo{author}{Jose~A. \surnamestart Galindo\surnameend}
  (\bibinfo{year}{2011}): \emph{\bibinfo{title}{Configuration of Multi Product
  Lines by Bridging Heterogeneous Variability Modeling Approaches}}.
\newblock In: {\sl \bibinfo{booktitle}{Proceedings of the 15th International
  Software Product Line Conference (SPLC 2011)}}, \bibinfo{address}{Munich,
  Germany}, \doi{10.1109/SPLC.2011.22}.

\bibitemdeclare{techreport}{KCH+1990Feature}
\bibitem{KCH+1990Feature}
\bibinfo{author}{Kyo~C. \surnamestart Kang\surnameend},
  \bibinfo{author}{Sholom~G. \surnamestart Cohen\surnameend},
  \bibinfo{author}{James~A. \surnamestart Hess\surnameend},
  \bibinfo{author}{William~E. \surnamestart Novak\surnameend} \&
  \bibinfo{author}{A.~Spencer \surnamestart Peterson\surnameend}
  (\bibinfo{year}{1990}): \emph{\bibinfo{title}{Feature Oriented Domain
  Analysis ({FODA}) Feasibility Study}}.
\newblock \bibinfo{type}{SEI Technical Report}
  \bibinfo{number}{CMU/SEI-90-TR-21, ADA 235785},
  \bibinfo{institution}{Software Engineering Institute}.
\newblock \urlprefix\url{http://www.sei.cmu.edu/reports/90tr021.pdf}.

\bibitemdeclare{techreport}{Krygiel1999Behind}
\bibitem{Krygiel1999Behind}
\bibinfo{author}{A.J. \surnamestart Krygiel\surnameend} (\bibinfo{year}{1999}):
  \emph{\bibinfo{title}{Behind the Wizard's Curtain. An Integration Environment
  for a System of Systems}}.
\newblock \bibinfo{type}{Technical Report}, \bibinfo{institution}{DTIC
  Document}.
\newblock \urlprefix\url{http://www.dodccrp.org/files/Krygiel_Wizards.pdf}.

\bibitemdeclare{inproceedings}{Lehman1996Laws}
\bibitem{Lehman1996Laws}
\bibinfo{author}{M.~M. \surnamestart Lehman\surnameend} (\bibinfo{year}{1996}):
  \emph{\bibinfo{title}{Laws of Software Evolution Revisited}}.
\newblock In \bibinfo{editor}{Carlo \surnamestart Montangero\surnameend},
  editor: {\sl \bibinfo{booktitle}{EWSPT}}, {\sl \bibinfo{series}{Lecture Notes
  in Computer Science}} \bibinfo{volume}{1149}, \bibinfo{publisher}{Springer},
  pp. \bibinfo{pages}{108--124}, \doi{10.1007/BFb0017737}.

\bibitemdeclare{article}{Lehman2001Rules}
\bibitem{Lehman2001Rules}
\bibinfo{author}{M.~M. \surnamestart Lehman\surnameend} \&
  \bibinfo{author}{Juan~F. \surnamestart Ramil\surnameend}
  (\bibinfo{year}{2001}): \emph{\bibinfo{title}{Rules and Tools for Software
  Evolution Planning and Management}}.
\newblock {\sl \bibinfo{journal}{Ann. Software Eng.}}
  \bibinfo{volume}{11}(\bibinfo{number}{1}), pp. \bibinfo{pages}{15--44},
  \doi{10.1023/A:1012535017876}.

\bibitemdeclare{article}{Maier1998Architecting}
\bibitem{Maier1998Architecting}
\bibinfo{author}{M.W. \surnamestart Maier\surnameend} (\bibinfo{year}{1998}):
  \emph{\bibinfo{title}{Architecting principles for systems-of-systems}}.
\newblock {\sl \bibinfo{journal}{Systems Engineering}}
  \bibinfo{volume}{1}(\bibinfo{number}{4}), pp. \bibinfo{pages}{267--284},
  \doi{10.1002/(SICI)1520-6858(1998)1:4<267::AID-SYS3>3.0.CO;2-D}.

\bibitemdeclare{inproceedings}{Matinlassi2004Comparison}
\bibitem{Matinlassi2004Comparison}
\bibinfo{author}{Mari \surnamestart Matinlassi\surnameend}
  (\bibinfo{year}{2004}): \emph{\bibinfo{title}{Comparison of Software Product
  Line Architecture Design Methods: COPA, FAST, FORM, KobrA and QADA}}.
\newblock In: {\sl \bibinfo{booktitle}{ICSE}}, pp. \bibinfo{pages}{127--136}.
\newblock
  \urlprefix\url{http://csdl.computer.org/comp/proceedings/icse/2004/2163/00/2%
1630127abs.htm}.

\bibitemdeclare{book}{Mens2008Software}
\bibitem{Mens2008Software}
\bibinfo{editor}{Tom \surnamestart Mens\surnameend} \& \bibinfo{editor}{Serge
  \surnamestart Demeyer\surnameend}, editors (\bibinfo{year}{2008}):
  \emph{\bibinfo{title}{Software Evolution}}.
\newblock \bibinfo{publisher}{Springer}, \doi{10.1007/978-3-540-76440-3}.

\bibitemdeclare{article}{Parnas1976Design}
\bibitem{Parnas1976Design}
\bibinfo{author}{D.~\surnamestart Parnas\surnameend} (\bibinfo{year}{1976}):
  \emph{\bibinfo{title}{On the Design and Development of Program Families}}.
\newblock {\sl \bibinfo{journal}{IEEE Transactions on Software Engineering}}
  \bibinfo{volume}{SE-2}(\bibinfo{number}{1}), pp. \bibinfo{pages}{1--9},
  \doi{10.1109/TSE.1976.233797}.

\bibitemdeclare{article}{Pleuss2012Visualization}
\bibitem{Pleuss2012Visualization}
\bibinfo{author}{Andreas \surnamestart Pleuss\surnameend} \&
  \bibinfo{author}{Goetz \surnamestart Botterweck\surnameend}
  (\bibinfo{year}{2012}): \emph{\bibinfo{title}{Visualization of variability
  and configuration options}}.
\newblock {\sl \bibinfo{journal}{International Journal on Software Tools for
  Technology Transfer (STTT)}}, pp. \bibinfo{pages}{1--14},
  \doi{10.1007/s10009-012-0252-z}.

\bibitemdeclare{article}{Pleuss2012Model}
\bibitem{Pleuss2012Model}
\bibinfo{author}{Andreas \surnamestart Pleuss\surnameend},
  \bibinfo{author}{Goetz \surnamestart Botterweck\surnameend},
  \bibinfo{author}{Deepak \surnamestart Dhungana\surnameend},
  \bibinfo{author}{Andreas \surnamestart Polzer\surnameend} \&
  \bibinfo{author}{Stefan \surnamestart Kowalewski\surnameend}
  (\bibinfo{year}{2012}): \emph{\bibinfo{title}{Model-driven Support for
  Product Line Evolution on Feature Level}}.
\newblock {\sl \bibinfo{journal}{Journal of Systems and Software (JSS) -
  Special Issue on Automated Software Evolution}}
  \bibinfo{volume}{85}(\bibinfo{number}{10}), pp. \bibinfo{pages}{2261--2274},
  \doi{10.1016/j.jss.2011.08.008}.

\bibitemdeclare{book}{PBL2005Software}
\bibitem{PBL2005Software}
\bibinfo{author}{Klaus \surnamestart Pohl\surnameend},
  \bibinfo{author}{G{\"u}nter \surnamestart B{\"o}ckle\surnameend} \&
  \bibinfo{author}{Frank \surnamestart van~der Linden\surnameend}
  (\bibinfo{year}{2005}): \emph{\bibinfo{title}{Software Product Line
  Engineering : Foundations, Principles, and Techniques}}.
\newblock \bibinfo{publisher}{Springer}, \bibinfo{address}{New York, NY}.
\newblock
  \urlprefix\url{http://www.springer.com/computer/swe/book/978-3-540-24372-4}.

\bibitemdeclare{inproceedings}{Reiser2006Managing}
\bibitem{Reiser2006Managing}
\bibinfo{author}{M.-O. \surnamestart Reiser\surnameend} \&
  \bibinfo{author}{M.~\surnamestart Weber\surnameend} (\bibinfo{year}{2006}):
  \emph{\bibinfo{title}{Managing Highly Complex Product Families with
  Multi-Level Feature Trees}}.
\newblock In: {\sl \bibinfo{booktitle}{Requirements Engineering, 14th IEEE
  International Conference}}, pp. \bibinfo{pages}{149 --158},
  \doi{10.1109/RE.2006.39}.

\bibitemdeclare{article}{SW1998Product}
\bibitem{SW1998Product}
\bibinfo{author}{Daniel \surnamestart Sabin\surnameend} \&
  \bibinfo{author}{Rainer \surnamestart Weigel\surnameend}
  (\bibinfo{year}{1998}): \emph{\bibinfo{title}{Product Configuration
  Frameworks - A Survey}}.
\newblock {\sl \bibinfo{journal}{IEEE Intelligent Systems and Applications}}
  \bibinfo{volume}{13}(\bibinfo{number}{4}), pp. \bibinfo{pages}{42--49},
  \doi{10.1109/5254.708432}.

\bibitemdeclare{article}{Sage2001systems}
\bibitem{Sage2001systems}
\bibinfo{author}{A.P. \surnamestart Sage\surnameend} \& \bibinfo{author}{C.D.
  \surnamestart Cuppan\surnameend} (\bibinfo{year}{2001}):
  \emph{\bibinfo{title}{On the systems engineering and management of systems of
  systems and federations of systems}}.
\newblock {\sl \bibinfo{journal}{Information Knowledge Systems Management}}
  \bibinfo{volume}{2}(\bibinfo{number}{4}), pp. \bibinfo{pages}{325--345}.
\newblock
  \urlprefix\url{http://iospress.metapress.com/content/wx6b5wft80k8p9a2/}.

\bibitemdeclare{inproceedings}{SS2000PuLSE-BEAT}
\bibitem{SS2000PuLSE-BEAT}
\bibinfo{author}{K.~\surnamestart Schmid\surnameend} \&
  \bibinfo{author}{M.~\surnamestart Schank\surnameend} (\bibinfo{year}{2000}):
  \emph{\bibinfo{title}{PuLSE-BEAT - A Decision Support Tool for Scoping
  Product Lines}}.
\newblock In: {\sl \bibinfo{booktitle}{Third International Workshop on Software
  Architectures for Product Families}}, pp. \bibinfo{pages}{64--74},
  \doi{10.1007/978-3-540-44542-5\_8}.

\bibitemdeclare{inproceedings}{SPB+2012Modeling}
\bibitem{SPB+2012Modeling}
\bibinfo{author}{Mathias \surnamestart Schubanz\surnameend},
  \bibinfo{author}{Andreas \surnamestart Pleuss\surnameend},
  \bibinfo{author}{Goetz \surnamestart Botterweck\surnameend} \&
  \bibinfo{author}{Claus \surnamestart Lewerentz\surnameend}
  (\bibinfo{year}{2012}): \emph{\bibinfo{title}{Modeling rationale over time to
  support product line evolution planning}}.
\newblock In: {\sl \bibinfo{booktitle}{Proceedings of the Sixth International
  Workshop on Variability Modeling of Software-Intensive Systems}},
  \bibinfo{series}{VaMoS '12}, \bibinfo{publisher}{ACM}, \bibinfo{address}{New
  York, NY, USA}, pp. \bibinfo{pages}{193--199}, \doi{10.1145/2110147.2110169}.

\bibitemdeclare{inproceedings}{Siegmund2012Predicting}
\bibitem{Siegmund2012Predicting}
\bibinfo{author}{Norbert \surnamestart Siegmund\surnameend},
  \bibinfo{author}{Sergiy~S. \surnamestart Kolesnikov\surnameend},
  \bibinfo{author}{Christian \surnamestart K{\"a}stner\surnameend},
  \bibinfo{author}{Sven \surnamestart Apel\surnameend}, \bibinfo{author}{Don
  \surnamestart Batory\surnameend}, \bibinfo{author}{Marko \surnamestart
  Rosenm{\"u}ller\surnameend} \& \bibinfo{author}{Gunter \surnamestart
  Saake\surnameend} (\bibinfo{year}{2012}): \emph{\bibinfo{title}{Predicting
  Performance via Automated Feature-Interaction Detection}}.
\newblock In: {\sl \bibinfo{booktitle}{Proceedings of International Conference
  on Software Engineering (ICSE)}}, \bibinfo{publisher}{IEEE}, pp.
  \bibinfo{pages}{167--177}, \doi{10.1109/ICSE.2012.6227196}.
\newblock
  \urlprefix\url{http://wwwiti.cs.uni-magdeburg.de/iti\_db/publikationen/ps/au%
to/SKK+12.pdf}.

\bibitemdeclare{inproceedings}{Siegmund2012Interoperability}
\bibitem{Siegmund2012Interoperability}
\bibinfo{author}{Norbert \surnamestart Siegmund\surnameend},
  \bibinfo{author}{Maik \surnamestart Mory\surnameend}, \bibinfo{author}{Janet
  \surnamestart Feigenspan\surnameend}, \bibinfo{author}{Gunter \surnamestart
  Saake\surnameend}, \bibinfo{author}{Mykhaylo \surnamestart
  Nykolaychuk\surnameend} \& \bibinfo{author}{Marco \surnamestart
  Schumann\surnameend} (\bibinfo{year}{2012}):
  \emph{\bibinfo{title}{Interoperability of Non-functional Requirements in
  Complex Systems}}.
\newblock In: {\sl \bibinfo{booktitle}{ICSE2012: International Workshop on
  Software Engineering for Embedded Systems}}, \bibinfo{publisher}{IEEE}, pp.
  \bibinfo{pages}{2--8}, \doi{10.1109/SEES.2012.6225487}.
\newblock
  \urlprefix\url{http://wwwiti.cs.uni-magdeburg.de/iti\_db/publikationen/ps/au%
to/SMF+12.pdf}.

\bibitemdeclare{article}{Siegmund2011SPL}
\bibitem{Siegmund2011SPL}
\bibinfo{author}{Norbert \surnamestart Siegmund\surnameend},
  \bibinfo{author}{Marko \surnamestart Rosenm{\"u}ller\surnameend},
  \bibinfo{author}{Martin \surnamestart Kuhlemann\surnameend},
  \bibinfo{author}{Christian \surnamestart K{\"a}stner\surnameend},
  \bibinfo{author}{Sven \surnamestart Apel\surnameend} \&
  \bibinfo{author}{Gunter \surnamestart Saake\surnameend}
  (\bibinfo{year}{2011}): \emph{\bibinfo{title}{SPL Conqueror: Toward
  Optimization of Non-functional Properties in Software Product Lines}}.
\newblock {\sl \bibinfo{journal}{Software Quality Journal}} \bibinfo{volume}{to
  appear}, \doi{10.1007/s11219-011-9152-9}.

\bibitemdeclare{article}{Thompson2003Structuring}
\bibitem{Thompson2003Structuring}
\bibinfo{author}{Jeffrey~M. \surnamestart Thompson\surnameend} \&
  \bibinfo{author}{Mats Per~Erik \surnamestart Heimdahl\surnameend}
  (\bibinfo{year}{2003}): \emph{\bibinfo{title}{Structuring product family
  requirements for n-dimensional and hierarchical product lines}}.
\newblock {\sl \bibinfo{journal}{Requir. Eng.}}
  \bibinfo{volume}{8}(\bibinfo{number}{1}), pp. \bibinfo{pages}{42--54},
  \doi{10.1007/s00766-003-0166-0}.

\bibitemdeclare{article}{White2009Selecting}
\bibitem{White2009Selecting}
\bibinfo{author}{Jules \surnamestart White\surnameend}, \bibinfo{author}{Brian
  \surnamestart Dougherty\surnameend} \& \bibinfo{author}{Douglas~C.
  \surnamestart Schmidt\surnameend} (\bibinfo{year}{2009}):
  \emph{\bibinfo{title}{Selecting highly optimal architectural feature sets
  with Filtered Cartesian Flattening}}.
\newblock {\sl \bibinfo{journal}{Journal of Systems and Software}}
  \bibinfo{volume}{82}(\bibinfo{number}{8}), pp. \bibinfo{pages}{1268--1284},
  \doi{10.1016/j.jss.2009.02.011}.

\end{thebibliography}

\end{document}